# Title: Discovery and spectroscopy of the young Jovian planet 51 Eri b with the Gemini Planet Imager

**Authors:**

B. Macintosh[1,2*], J. R. Graham[3], T. Barman[4], R. J. De Rosa[3], Q. Konopacky[5], M. S. Marley[6], C. Marois[7,8], E. L. Nielsen[9,1], L. Pueyo[10], A. Rajan[11], J. Rameau[12], D. Saumon[13], J. J. Wang[3], J. Patience[11], M. Ammons[2], P. Arriaga[14], E. Artigau[12], S. Beckwith[3], J. Brewster[9], S. Bruzzone[15], J. Bulger[11,16], B. Burningham[6,17], A. S. Burrows[18], C. Chen[10], E. Chiang[3], J. K. Chilcote[19], R. I. Dawson[3], R. Dong[3], R. Doyon[12], Z. H. Draper[8,7], G. Duchêne[3,20], T. M. Esposito[14], D. Fabrycky[21], M. P. Fitzgerald[14], K. B. Follette[1], J. J. Fortney[22], B. Gerard[8,7], S. Goodsell[23,24], A. Z. Greenbaum[25,10], P. Hibon[24], S. Hinkley[26], T. H. Cotten[27], L.-W. Hung[14], P. Ingraham[28], M. Johnson-Groh[8,7], P. Kalas[3,9], D. Lafreniere[12], J. E. Larkin[14], J. Lee[27], M. Line[22], D. Long[10], J. Maire[19], F. Marchis[9], B. C. Matthews[7,8], C. E. Max[22], S. Metchev[15,29], M. A. Millar-Blanchaer[30], T. Mittal[3], C. V. Morley[22], K. M. Morzinski[31], R. Murray-Clay[32], R. Oppenheimer[33], D. W. Palmer[2], R. Patel[29], M. D. Perrin[10], L. A. Poyneer[2], R. R. Rafikov[18], F. T. Rantakyrö[24], E. L. Rice[34,33], P. Rojo[35], A. R. Rudy[22], J.-B. Ruffio[1,9], M. T. Ruiz[35], N. Sadakuni[36,24], L. Saddlemyer[7], M. Salama[3], D. Savransky[37], A. C. Schneider[38], A. Sivaramakrishnan[10], I. Song[27], R. Soummer[10], S. Thomas[28], G. Vasisht[39], J. K. Wallace[39], K. Ward-Duong[11], S. J. Wiktorowicz[22], S. G. Wolff[25,10], B. Zuckerman[14]

**Affiliations:**

[1]Kavli Institute for Particle Astrophysics and Cosmology, Stanford University, Stanford, CA 94305, USA.
[2]Lawrence Livermore National Laboratory, 7000 East Avenue, Livermore, CA 94040, USA.
[3]Department of Astronomy, University of California–Berkeley, Berkeley, CA, 94720, USA.
[4]Lunar and Planetary Laboratory, University of Arizona, Tucson, AZ 85721, USA.
[5]Center for Astrophysics and Space Sciences, University of California–San Diego, 9500 Gilman Drive, La Jolla, CA 92093, USA.
[6]NASA Ames Research Center, MS 245-3, Moffett Field, CA 94035, USA.
[7]National Research Council of Canada, Herzberg Institute of Astrophysics, 5071 West Saanich Road, Victoria, BC V9E 2E7, Canada.
[8]University of Victoria, 3800 Finnerty Road, Victoria, BC V8P 5C2, Canada.
[9]Search for Extraterrestrial Intelligence Institute, Carl Sagan Center, 189 Bernardo Avenue, Mountain View, CA 94043, USA.
[10]Space Telescope Science Institute, 3700 San Martin Drive, Baltimore, MD 21218, USA.
[11]School of Earth and Space Exploration, Arizona State University, PO Box 871404, Tempe, AZ 85287, USA.
[12]Institut de Recherche sur les Exoplanètes, Départment de Physique, Université de Montréal, Montréal, QC H3C 3J7, Canada.
[13]Los Alamos National Laboratory, P.O. Box 1663, MS F663, Los Alamos, NM 87545, USA.
[14]Department of Physics and Astronomy, University of California–Los Angeles, 430 Portola Plaza, Los Angeles, CA 90095, USA.
[15]Department of Physics and Astronomy, The University of Western Ontario, London, ON N6A 3K7, Canada.
[16]Subaru Telescope, 650 North A'ohoku Place, Hilo, HI 96720, USA.
[17]Science and Technology Research Institute, University of Hertfordshire, Hatfield AL10 9AB, UK.
[18]Department of Astrophysical Sciences, Princeton University, Princeton, NJ 08544, USA.
[19]Dunlap Institute for Astronomy and Astrophysics, University of Toronto, 50 St. George Street, Toronto, ON M5S 3H4, Canada.
[20]Institut de Planétologie et d'Astrophysique de Grenoble, Université Grenoble Alpes, Centre National de la Recherche Scientifique, 38000 Grenoble, France.

[21]Department of Astronomy and Astrophysics, University of Chicago, 5640 South Ellis Avenue, Chicago, IL 60637, USA.
[22]Department of Astronomy and Astrophysics, University of California–Santa Cruz, Santa Cruz, CA 95064, USA.
[23]Durham University, Stockton Road, Durham DH1, UK.
[24]Gemini Observatory, Casilla 603, La Serena, Chile.
[25]Department of Physics and Astronomy, Johns Hopkins University, 3600 North Charles Street, Baltimore MD, 21218, USA.
[26]University of Exeter, Astrophysics Group, Physics Building, Stocker Road, Exeter EX4 4QL, UK.
[27]Department of Physics and Astronomy, University of Georgia, Athens, GA 30602, USA.
[28]Large Synoptic Survey Telescope, 950 North Cherry Avenue, Tucson, AZ 85719, USA.
[29]Department of Physics and Astronomy, Stony Brook University, 100 Nicolls Road, Stony Brook, NY 11794-3800, USA.
[30]Department of Astronomy and Astrophysics, University of Toronto, Toronto, ON M5S 3H4, Canada.
[31]Steward Observatory, 933 North Cherry Avenue, University of Arizona, Tucson, AZ 85721, USA.
[32]Department of Physics, University of California–Santa Barbara, Broida Hall, Santa Barbara, CA 93106-9530, USA.
[33]American Museum of Natural History, New York, NY 10024, USA.
[34]Department of Engineering Science and Physics, College of Staten Island, City University of New York, Staten Island, NY 10314, USA.
[35]Departamento de Astronomía, Universidad de Chile, Camino El Observatorio 1515, Casilla 36-D, Las Condes, Santiago, Chile.
[36]Stratospheric Observatory for Infrared Astronomy, Universities Space Research Association, NASA/Armstrong Flight Research Center, 2825 East Avenue P, Palmdale, CA 93550, USA.
[37]Sibley School of Mechanical and Aerospace Engineering, Cornell University, Ithaca, NY 14853, USA.
[38]Physics and Astronomy, University of Toledo, 2801 West Bancroft Street, Toledo, OH 43606, USA.
[39]Jet Propulsion Laboratory, California Institute of Technology, 4800 Oak Grove Drive, Pasadena, CA 91109, USA.

*Correspondence to: bmacintosh@stanford.edu

**Abstract**: Directly detecting thermal emission from young extrasolar planets allows measurement of their atmospheric composition and luminosity, which is influenced by their formation mechanism. Using the Gemini Planet Imager, we discovered a planet orbiting the ~20 Myr-old star 51 Eridani at a projected separation of 13 astronomical units. Near-infrared observations show a spectrum with strong methane and water vapor absorption. Modeling of the spectra and photometry yields a luminosity of $L/L_{\odot}$=1.6-4.0 × $10^{-6}$ and an effective temperature of 600-750 K. For this age and luminosity, "hot-start" formation models indicate a mass twice that of Jupiter. This planet also has a sufficiently low luminosity to be consistent with the "cold-start" core accretion process that may have formed Jupiter.

**One Sentence Summary:** Using an advanced adaptive optics coronagraph we have discovered and spectrally characterized a young 2-12 Jupiter mass planet that shows strong methane spectral signatures and could have formed through low-entropy core accretion.

Several young self-luminous extrasolar planets have been directly imaged (*1–8*) at infrared wavelengths. The currently known directly imaged planets are massive (estimated 5-13 $M_J$) and at large separations (9-650 AU) from their host star, compared to our solar system. Photometry and spectroscopy probe the atmospheres of these young Jupiter, providing hints about their formation. Several unexpected results have emerged. The near-infrared colors of these planets are mostly red, indicating cloudy atmospheres reminiscent of brown dwarfs of spectral type L. Methane absorption features are prominent in the near-infrared spectra of T dwarfs (Teff <1100K), as well as the giant planets of our solar system, but so far weak or absent in the directly imaged exoplanets (*4*, *9–11*). Most young planets appear to be methane-free even at temperatures where equivalent brown dwarfs show methane, suggesting non-equilibrium chemistry and persistent clouds that are likely age and mass dependent (*1*, *12–15*).

In spite of uncertainties in their atmospheric properties, the luminosities of these planets are well constrained. Luminosity is a function of age, mass and initial conditions (*16*, *17*) and hence can provide insights into a planet's formation. Rapid formation, e.g., through global disk instabilities acting on a dynamical timescale, yields high-entropy planets that are bright at young ages – referred to as "hot-start". Alternatively, two-stage formation, first of a dense solid core followed by gas accretion through a shock, as likely in the case of Jupiter, can produce a range of states including lower-entropy planets that are cooler, and slightly smaller in radius ("cold start"). The young directly imaged planets are almost all too bright for "cold-start" except for very specific accretion shock properties, though their formation is also difficult to explain through global instability which should operate preferentially at higher masses and large semi-major axis separation (*18*, *19*). These planets are also close to the limit of sensitivity for first-generation large-telescope adaptive optics (AO) systems. The goal of the latest generation of surveys using dedicated high-contrast adaptive optics coronagraphs (*20–23*) such as the Gemini Planet Imager (GPI) and its counterparts is to expand this sample to closer separations, lower masses and temperatures, a crucial empirical step towards investigating these modes of formation.

The Gemini Planet Imager Exoplanet Survey (GPIES) is targeting 600 young, nearby stars using the GPI instrument. The star 51 Eridani (51 Eri) was chosen as an early target for the survey due to its youth and proximity. Stellar properties are given in Table 1. The star has weak mid- and far-IR excess emission indicating low-mass inner (5.5 AU) and outer (82 AU) dust belts(*24*, *25*). It also has two distant (~2000 AU) stellar companions—the 6 AU separation M-dwarf binary GJ3305AB (*26*). 51 Eri and GJ 3305 were classified in 2001 as members of the β Pictoris moving group (*27*) and subsequent measurements (*28*) support this identification. The estimated age of the β Pictoris moving group in the literature ranges from 12 to 23 Myr (*27*, *29–32*). Giving strong weight to the group's lithium depletion boundary age, we adopt an age of 20±6 Myr for all four components of the 51 Eri system (*28*).

We observed 51 Eri in *H* band (1.65 μm) in December 2014 as the 44$^{th}$ target in the GPIES campaign. GPI observations produce spectroscopic cubes with spectral resolving power of 45 over the entire field of view. A companion, designated 51 Eri b, was apparent after point spread function subtraction. The planet is located at a projected separation of 13 AU and showed distinctive strong methane and water vapor absorption (Fig. 1 and Fig. 2). We observed 51 Eri in January 2015 to broaden the wavelength coverage using GPI (*J* band; 1.25 μm), and NIRC2/W. M. Keck 2 (*Lp*; 3.8 μm) and recovered the planet in both observations. The observed spectrum is highly similar to a field brown dwarf of spectral type T4.5-T6 (Fig. 2). The *J*-band spectrum confirmed methane absorption at this wavelength and the extremely red *H-Lp* color is also similar to other cool, low-mass objects (Fig. 3). The signal-to-noise at *J*-band is inferior to that at *H*, and extraction introduces additional systematic effects. The *J*-band detection is reliable (> 6-σ), but the fluxes in individual spectral channels are less certain. However, the methane feature is robustly detected at both bands (*28*).

Demonstrating common proper motion (e.g. *34*) or showing that the probability of a foreground or background contaminant is extremely low establishes the nature of directly imaged planets. The interval between the December 2014 and January 2015 observations is too brief given our astrometric accuracy (*28*) to show that 51 Eri b and 51 Eri share proper motion and parallax. However, non-detection of 51 Eri b in archival data from 2003 (*28*) excludes a stationary background source and requires proper motion within ~0.1 arcsec/yr of 51 Eri. The strong methane absorption seen in 51 Eri b is found only in T-type or later brown dwarfs. We determined the probability of finding a T dwarf in our field by merging the observed T-dwarf luminosity functions (*27-28*) and adopting the spectral types and absolute magnitudes for T dwarfs (*34*) from which we calculate a false alarm rate of $1.72\times10^{-7}$ methane objects (i.e., types T0-T8.5) per GPI field (> 5- σ). The proper motion constraint eliminates a further 66% of likely background T dwarf proper motions. The total false alarm probability after observing 44 targets is that for a T spectrum object to appear in 44 Bernoulli trials, given by the binomial distribution, yielding a final probability of $2.4\times10^{-6}$. While the occurrence rates of planetary companions is not known with precision, the detection of planetary objects such as β Pic b and HR8799 e at similar physical separations to 51 Eri b indicate that the rate is $>10^{-3}$ per star. Hence, with the high-quality spectrum available to us, it is vastly more likely that 51 Eri b is a bound planetary companion than a chance alignment.

We use planetary atmosphere and evolution models to estimate the properties of 51 Eri b. We first fit the observed *J* and *H* band spectra using standard cloud-free equilibrium-chemistry models, constrained to have radii for a given mass as given by evolutionary tracks, similar to those in (*35*). This constrained fit gives an effective temperature of 750K, with a radius (0.76 $R_J$) and surface gravity similar to an old (10 Gyr), high-mass brown dwarf. A similar though less extreme result – small radii and hence high masses and old ages – is found in several model fits to the HR8799 observations (*13, 15 16*), even though high masses are clearly excluded by dynamical stability considerations (e.g. *32*). This model was not constrained to fit the *Lp* observation but does within 1.6σ.

We next fit a model to the *JH* spectra and *Lp* photometry using a linear combination of cloudy and cloud-free surfaces and non-equilibrium chemistry and allowed the planet's radius to vary independently of the radii given by evolutionary tracks. Models of this type generally produce reasonable fits to other directly imaged planets (*11*, *12*, *15*, *38–40*). This model produces a slightly lower effective temperature. The spectral shape and colors only weakly constrain gravity but do favor lower masses, while the radius (~1 $R_{Jup}$) is consistent with evolutionary tracks given the age of the system. Table 2 summarizes the results of the modeling. With the spectral and atmospheric uncertainties, a wide range of other models (including temperatures as high as ~1000K) are also broadly consistent with the observations. The low temperature is supported by the presence of strong methane absorption that is not seen in other planets of similar age.

The luminosity of $\log(L/L_\odot)$ of -5.4 to -5.8 is similar in all models regardless of temperature or clouds. Combined with the age, that luminosity can be used to estimate the mass of the planet. For a hot-start model, this corresponds to a mass of ~2 $(t/20\mathrm{My})^{0.65}(L/2\times10^{-6})^{0.54}$ $M_{Jup}$ – the lowest-mass self-luminous planet directly imaged to date. 51 Eri b, unlike other young (<100 Myr) planetary-mass companions, has a low enough luminosity low to be consistent with cold-start core accretion scenarios. In cold-start evolution, luminosity at 20 Myr age is nearly independent of mass, so the mass of 51 Eri b would be between 2 and 12 $M_{Jup}$.

51 Eri b and the GJ 3305 binary form a hierarchical triple configuration (*28*), but the companion pair is far enough away that the planet is expected to be dynamically stable in its current orbit (*38*). Moreover, the young age of the system suggests that while long-term dynamical effects such as secular Lidov-Kozai oscillations might have altered the planet's eccentricity and inclination, it is unlikely they have had time to produce the extreme eccentricities required for tidal friction to alter the planet's semi-major axis(*41*). The formation of a ~2 $M_{Jup}$ planet at an orbital distance of ~15 AU around a Sun-like star can be explained by modest extensions to the core accretion theory. Early versions of the theory (*42*) found that accretion of the core at larger orbital distances is in danger of taking too long, failing to capture the natal gas before it dissipates. 51 Eri b is close enough to the star that this may be less of a problem, and the addition of migration (*43*) or pebbles that experience gas drag (*44*) also help overcome this timescale difficulty.

The transition from L-type to T-type planets appears to occur over a narrow range of temperatures between the ~1000K HR8799b or PSO J318.5-22(*44*) and the 700K 51 Eri. Direct determination of the object's mass – either through spectral surface gravity indicators, or reflex astrometry of the primary star – could determine whether it formed through hot- or cold-start processes; 51 Eri b provides an opportunity to study in detail a planet that is still influenced by its formation initial conditions. With a methane-dominated spectrum, low luminosity and potentially low-entropy start, 51 Eri b is a bridge from wider-orbit, hotter and more massive planets to Jupiter-like scales.

## Acknowledgments:

Based on observations obtained at the Gemini Observatory, which is operated by the Association of Universities for Research in Astronomy, Inc., under a cooperative agreement with the NSF on behalf of the Gemini partnership: the National Science Foundation (United States), the National Research Council (Canada), CONICYT (Chile), the Australian Research Council (Australia), Ministério da Ciência, Tecnologia e Inovação (Brazil) and Ministerio de Ciencia, Tecnología e Innovación Productiva (Argentina). Supported by grants from the National Science Foundation AST-1411868 (BM, KF, JP, AR), AST-0909188 and 1313718 (JRG, PK, RDR, JW), AST-1313718 (MPF and GD) and AST-1405505. Supported by grants from NASA NNX14AJ80G (BM, FM, EN, MP), NNH15AZ591 (DS, MM), NNX15AD95G (JRG, PK), NNX11AD21G (JRG, PK) and NNH11ZDA001N (SM, RP). JR, RD and DL acknowledge support from the Fonds de recherche du Quebec. AZG acknowledges NSF fellowship DGE-123825. K.W.-D. acknowledges NSF fellowship DGE-1311230. LWH acknowledges NSF fellowship DGE-1144087Portions of this work were performed under the auspices of the U.S. Department of Energy by Lawrence Livermore National Laboratory under Contract DE-AC52-07NA27344. GPI data are archived at the Gemini Science Archive, http://www.cadc-ccda.hia-iha.nrc-cnrc.gc.ca/en/gsa/

**Fig. 1. Images of 51 Eri and 51 Eri b (indicated by arrow) after point spread function (PSF) subtraction.** (a) *H*-band GPI image from December 2014. (b) *J*-band GPI image from January 2015. (c) *Lp*-band Keck image using the NIRC2 camera from January 2015.

**Fig. 2. 51 Eri b J and H band spectrum from GPI data after PSF subtraction.** Strong methane absorption, similar to Jupiter, is readily apparent. Top: The hotter young planetary object 2M1207b and a high-mass field T6 brown dwarf from the SpeX library (*45*) are overplotted. Bottom: Observed *J* and *H* spectrum and *Lp* photometry with two model fits overlaid, a young low-mass partly-cloudy object (TB-700K) and a higher-mass cloud-free object (SM-750K). Note that the main source of error in the extracted spectrum is residual speckle artifacts, so errors in neighboring spectral channels are strongly correlated; error estimation is discussed in (28)

**Fig. 3. Color-magnitude diagram of brown dwarfs (grey and black) and planetary-mass objects (colors)**. 51 Eri b is indicated with a red star, distinct from most other planets in the methane-dominated T-dwarf region of the diagram. The *Lp* field brown dwarf photometry is taken from (47, 48) or converted from WISE W1 (49) using an *Lp* vs. W1 linear fit. Parallaxes are available for all objects plotted (48).

| 51 Eridani | |
|---|---|
| Spectral Type | F0IV |
| Mass ($M_{\odot}$) | 1.75 ± 0.05 |
| Luminosity ($L_{\odot}$) | 7.1 ± 1 |
| Distance (pc) | 29.4 ± 0.3 |
| Proper Motion (mas/yr east, mas/yr north) | [44.22 ± 0.34, -64.39 ± 0.27] (*46*) |
| Age (Myr) | 20 ± 6 |
| Metallicity (M/H) | -0.027 (*47*) |
| *J*, *H*, *K*s, *L*p (mag.) | 4.74±0.04, 4.77±0.08, 4.54±0.02, 4.52±0.21 |
| $F_{dust}/F_{bol}$ | ~$10^{-6}$ |
| 51 Eri b | |
| Projected separation (mas) | 449±7 (31 January 2015; *33*) |
| Projected separation (AU) | 13.2±0.2 (31 January 2015) |
| $M_J$ | 16.75±0.40 |
| $M_H$ | 16.86±0.21 |
| $M_{Lp}$ | 13.82±0.27 |

**Table 1.** Properties of 51 Eridani A and 51 Eridani b

| | Cloud-free equilibrium model SM-750K | Partial-cloud model TB-700K |
|---|---|---|
| $M_J$ | 16.82 | 16.64 |
| $M_H$ | 17.02 | 16.88 |
| $M_L$ | 14.3 | 13.96 |
| $T_{eff}$ (K) | 750 | 700 |
| $R$ ($R_J$) | 0.76 | 1 |
| log $L/L_{\odot}$ | -5.8 | -5.6 |
| log($g$) | 5.5 | 3.5 |
| Age (Myr) | 10,000 | 20 (assumed to match stellar age) |
| Mass ($M_J$) | 67 | 2 (from luminosity, assuming high-entropy start) |

**Table 2.** Modeling results for 51 Eri b

**Supplementary Materials:**

Materials and Methods

Figures S1-S3

Tables S1-S3

References *45 - 93*

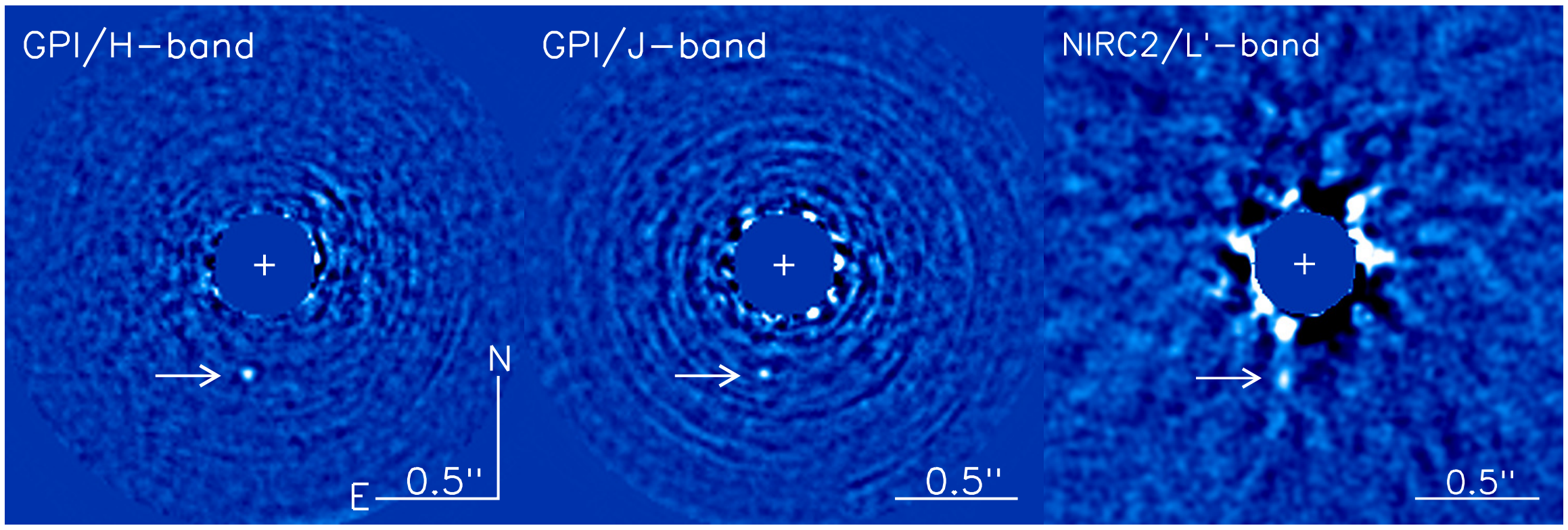

Fig. 1

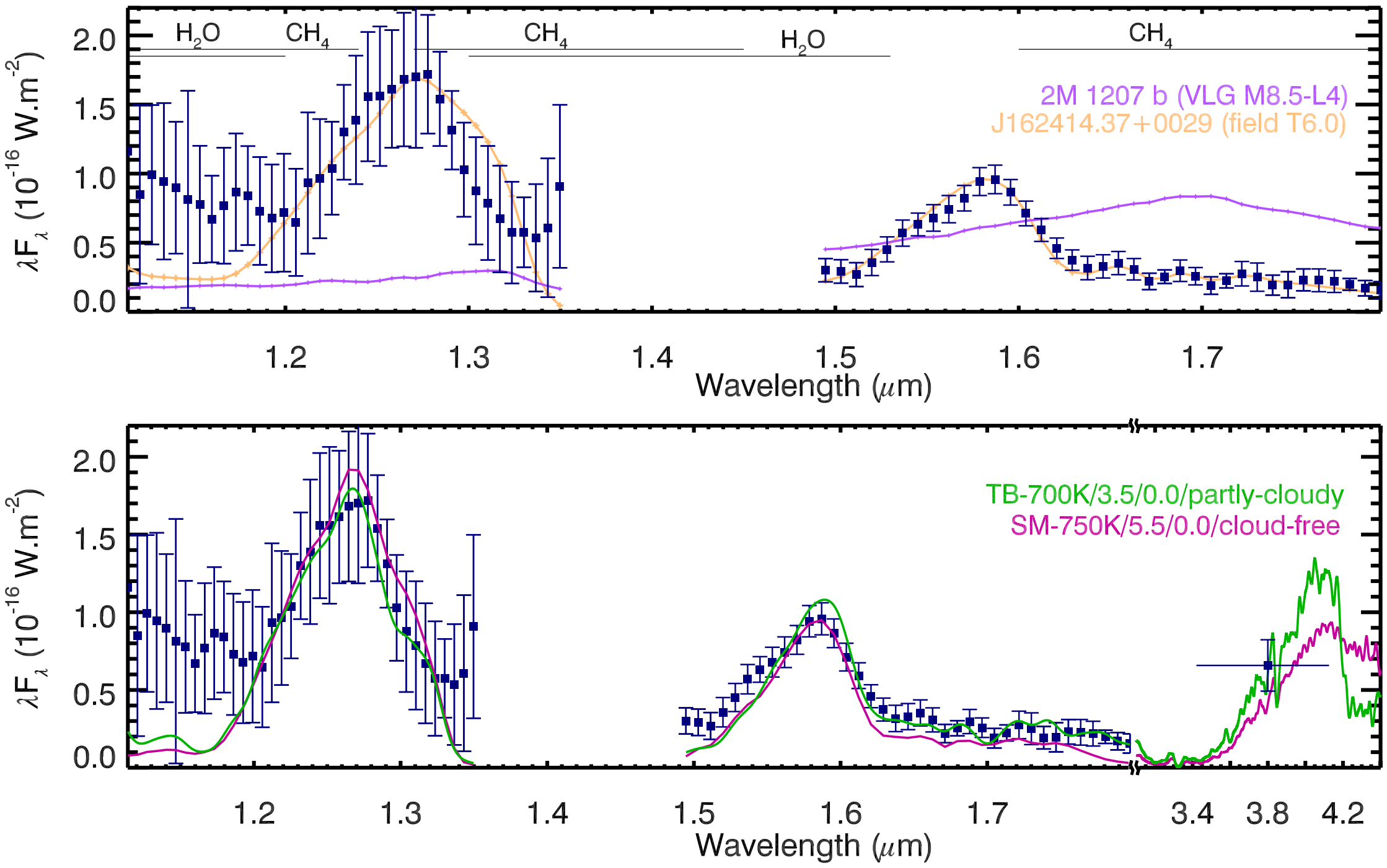

Fig. 2

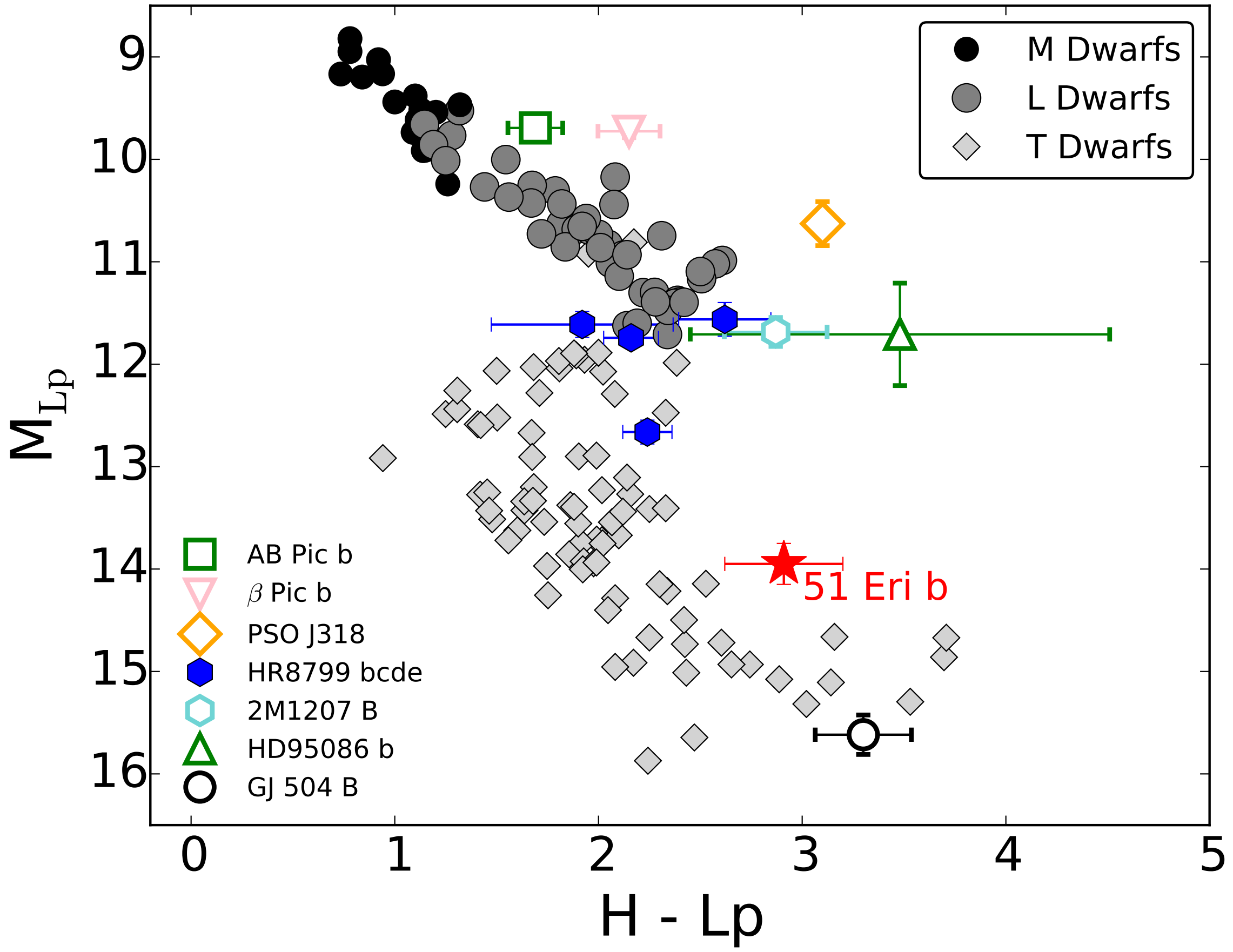

Fig. 3

**Supplementary Materials:**

**The Gemini Planet Imager**: The Gemini Planet Imager (GPI, *49*) is a facility-class, high-order, adaptive optics, imaging spectrograph designed to search for and characterize gas giant planets that orbit nearby young stars. At present the instrument is mounted on the Gemini South 8.1-meter telescope on Cerro Pachon in Chile. Detection and characterization of exoplanets by the direct imaging technique is challenging due to the typical small angular separation between a star and a planet, only a few times the diffraction width in the near-infrared for current 8–10 m class telescopes, and due to the large contrast ratio between a star and a planet. The GPI instrument detects exoplanets by their thermal emission. Because GPI-imaged planets are young, typically less than 100 Myr old, their atmospheres are still warm (~1000 K) from the accumulation and subsequent slow release of gravitational potential energy that was stored as the planets formed. Planets do not generate energy by nuclear fusion so they slowly cool with time, eventually becoming too faint to be detected by GPI; thus, GPI observes young stars in order to detect self-luminous planets. GPI's science instrument is a near-infrared integral field spectrograph (IFS, *50*).The spatial elements have a resolution of 0.014 arcseconds, and for each spatial pixel, GPI obtains a near-infrared spectrum with $\Delta\lambda/\lambda$~40. GPI observes in the near-infrared (0.9–2.2 μm) and a single observation covers one of the standard astronomical near-infrared bands, *Y, J, H,* or the blue or red half of the *K*-band. For more technical details on the instrument, see Macintosh et al. (*48*).

**The galactic space motion and age of the 51 Eri system**: Based on the GPI observations, 51 Eri and its planet 51 Eri b have a projected sky-plane separation of ~13.2 AU. The close binary GJ 3305 has a projected separation of ~6 AU (*26*), while 51 Eri and GJ 3305 have a projected separation of ~1960 AU (*50*). While the presence of a stellar companion with semi-major axis of ≲1000 AU may suppress the formation and/or continued existence of a planet like 51 Eri b around an A or F-type star (*51*), the GJ 3305 binary is at a significantly wider separation and has probably not had a significant impact on the formation history of 51 Eri b.

The brightest star in the system, 51 Eri is of spectral type F0IV (*52*) and has a visual magnitude of 5.2. The two components of GJ 3305 are of spectral type early and mid-M (e.g. *54*, *55*). An analysis of the dynamics of the GJ 3305 system has been done in Delorme et al. (*55*) and a refinement will appear in De Rosa et al. (2015, in prep.) and Nielsen et al. (2015, in prep).

To estimate a mass for 51 Eri b, we must determine its age. Various techniques for determining ages of youthful stars are considered in Zuckerman & Song (*56*). These techniques can be quite different for an early-type star such as 51 Eri and a late-type star like GJ 3305. If GJ 3305 can be shown convincingly to be a companion of 51 Eri, then estimated ages of both can be used to pin down the age of the planet 51 Eri b. 51 Eri and GJ 3305 were first classified as members of the β Pictoris moving group (BPMG, *28*).

Stars in the BPMG all have approximately the same 3-dimensional space velocities (U,V,W) in the Milky Way galaxy. The UVW of the BPMG is -10.1±2.1, -15.9±0.8, -9.2±1.0 km/s (*57*) where the quoted uncertainties are the dispersions among the velocities of the various members of the BPMG. To calculate the UVW of 51 Eri, we combine the parallax (33.98±0.34 mas) and proper motion in Right Ascension (44.22±0.34 mas/yr) and Declination (-64.39±0.27 mas/yr) given in van Leeuwen (*46*) with the stellar radial velocity. Two measurements of the radial velocity of 51 Eri appear in the literature: 21±5 km/s (*58*) and 12.6±3.5 km/s (*59*). Due to the difference between these two values, we measured the radial velocity of 51 Eri on 10 February 2015 (UT) with the Echelle Spectrograph Imager (ESI; *62*) on the Keck 2 telescope at the Mauna Kea Observatory. The ESI has a spectral resolving power of 13,000 and the measured signal-to-noise of the spectra were at least 100, so central wavelengths of various lines of Ca I and of Hα (6563 Å) could be measured very accurately. The dominant errors were systematic as judged from measurement of the radial velocity standard HIP 40843 (32.7±0.1 km/s; *63*). From our ESI observations, we measured a radial velocity for 51 Eri of 17.3±3 km/s, yielding a UVW space motion of -11.0, -15.2, -8.2 km/s, consistent with known BPMG members (*57*). In addition to the kinematics, the location of 51 Eri on an $M_V$ vs B-V color- magnitude diagram relative to evolutionary models is consistent with youth. A Bayeseian analysis using the BANYAN II code (*62*) gives a 97% probability of BPMG membership. A similarly high (99%) membership probability in BPMG is inferred for GJ3305AB assuming the same parallax as 51 Eri and a radial velocity of 20.63+/-0.46 km/s (*63*).

Additional evidence of BPMG membership can be provided by establishment of common space motion of 51 Eri and GJ 3305 in combination with demonstration of the youth of the GJ 3305 binary. The discussion in Feigelson et al. (*50*) demonstrates conclusively that 51 Eri and GJ 3305 constitute a bound system. The youth of GJ 3305 is supported by measurements of its X-ray luminosity and lithium abundance (*26*, *50*). In particular, the *V-K* color of GJ 3305, and its lithium equivalent width as measured by Feigelson et al. (*50*), is consistent with placement of GJ 3305 at one edge of the lithium depletion region for stars in the BPMG, as indicated in Figure 2 in Binks & Jeffries (*64*).

Various measurements of the radial velocity of the GJ 3005 binary system appear in the published literature. Bailey et al (*63*) report a mean velocity of 20.62 km/s with a standard deviation of 0.5 km/s over 5 epochs. Our analysis indicates a likely system radial velocity of 21.2 km/s with an uncertainty of about 0.5 km/s (a detailed discussion will appear in Nielsen et al. 2015, in prep). Assuming the same proper motion and distance as 51 Eri, and a radial velocity of 21.2±0.5 km/s, the UVW of GJ 3305 is -14.2,-16.2,-10.1 km/s. The V and W components agree well with the UVW of the BPMG while the U component differs by twice the dispersion of other members of the BPMG as given above.

Significant differences for the estimated age of the BPMG appear in the literature, ranging from 12 to 23 Myr (*29*, *30*, *32*, *57*) An age of 13 Myr for 51 Eri, independent from possible membership in the BPMG, was deduced from measurements with the CHARA optical interferometer on Mount Wilson (*29*). However, M. Simon subsequently uncovered some problems with the CHARA measurements, and now favors an age closer to 20 Myr than 13 Myr.

In summary, 51 Eri and GJ 3305 comprise a bound youthful multiple system with an age consistent with that of the BPMG. The only existing measurement for either star that might be at odds with the association of 51 Eri and the BPMG is the U component of the Galactic space motion of GJ 3305. Giving strong weight to the lithium depletion boundary age of the BPMG, we adopt an age of 20±6 Myr for all four components of the 51 Eri system.

**51 Eri Observations**: 51 Eri was first observed with GPI on 18 December 2014 UT as part of the GPI Exoplanet Survey (GPIES, program GS-2014B-Q-500). The sequence was acquired following the Angular Differential Imaging (ADI) observing technique (*65*). A total of thirty eight *H*-band images of 60-s integration were acquired for a total 2280-s on source exposure and 23.8 degrees field-of-view (FoV) rotation. The star was positioned at the center of the focal plane mask of the instrument to maximize starlight suppression. The observing conditions were excellent. Additional GPI sequences to characterize the system were then acquired on 29 January (*J* band), 30 January (*J* band) and 31 January (*H* band), 2015 (see Table S1 for more details about the observing sequences).

In addition to the GPI data, a 62-image ADI sequence of 54-s exposures was acquired using the NIRC2 narrow-field camera (P.I. K. Matthews) and the facility Keck adaptive optics system at the W.M. Keck 2 telescope on 1 February 2015, at *Lp* band, for a total of 3348 s on source exposure. The conditions were excellent and a total of 44.6 degrees FoV rotation was obtained. The semi-transparent 400 mas focal plane mask was used to block the starlight. A set of unsaturated, off focal plane mask images were acquired before the sequence to calibrate the star-to-planet flux ratio, and background images were acquired before and after the 51 Eri sequence.

**Data Reduction**: The initial GPI data reduction was done using an automated data processing system designed for the GPIES campaign (*66*). For each 2-D raw image, it performs dark subtraction and bad pixel removal. Microspectra are extracted with an argon lamp wavelength calibration taken immediately before the observations, producing a final 3-D spectral cube with 37 wavelength slices. The 3-D spectral cubes were then fixed again for bad pixels and corrected for distortion. The positions of the four fiducial diffraction spots, which we term satellite spots, were then measured for every frame. The satellite spots are used to determine the location of the occulted star, allowing us to align all the frames to a common center. The flux of the satellite spots is directly proportional to the flux of the occulted star, giving a relative flux calibration within each wavelength slice. Spectral cubes with evidence of poor data quality were rejected manually.

The NIRC2 *Lp* data was reduced using a custom IDL script. A dark current image was first subtracted from all images to remove electronic noises. Sky background acquired on an empty part of the sky near 51 Eri was subtracted from all the science images to remove the thermal background emission. The bad pixels were then corrected using the average of surrounding pixels, and the images were flat fielded using a *Kp*-band flat image having the same 400 mas focal plane mask. Optical distortions were corrected using the Yelda et al. (*67*) NIRC2 distortion solution. Conditions were photometric with less than 5% flux variations during the entire sequence (estimated from the stellar PSF core seen through the NIRC2 semi-transparent focal plane mask). The images were then flux normalized so that the stellar PSF core shows the same integrated flux in a 2.5 $\lambda/D$ diameter aperture for all images, and then registered at the image center using the stellar diffraction core and a cross-correlation algorithm.

**Stellar Halo and Speckle Subtraction**: Since all GPI spectral mode sequences are acquired with the ADI and the Spectral Difference Imaging (SDI) (*68*, *69*) techniques, it is possible to separate out astrophysical sources from the stellar halo using advanced post-processing algorithms. Following each observation sequence, the GPI automated data processing system produced preliminary reduction with the stellar halo subtracted. The automatic processing system uses the Principal Component Analysis (PCA/KLIP, *70*, *71*) approach to model and subtract out the stellar contribution. A Python implementation of the KLIP algorithm automatically processed the data as they were being taken to allow us to rapidly identify potential sources for follow-up.

Further processing and characterization were done outside the automated system using three custom independent pipelines (1, 2, and 3). Each data cube image slice was first high-pass filtered (one pipeline (1) using a Fourier filter to remove one fourth of the lowest spatial frequencies, another (2) using a Gaussian filter with FWHM of ~11 pixels, and the other (3) using a simple unsharp masking with a 11×11 pixel box window) to remove low-frequency components of the stellar halo. The slices were then scaled to align speckles as a function of wavelength using the mean separation of the four spots (if using SDI (1) or for all subtractions (3, 2)). We also normalized each slice by the mean flux of the four spots to remove stellar flux, atmospheric, and instrument throughput variations as a function of wavelength. The three scripts then perform the stellar halo subtraction. Each pipeline can be executed in two modes, one for finding planets, using a combination of SDI and ADI, and another to perform characterization, with ADI only.

Pipeline 1 has been adapted from Lagrange et al. (*3*) and Chauvin et al. (*72*). The scaled data cubes are first processed with the SDI step. For each spectral cube, a reference stellar halo image was built by taking the median of the rescaled and normalized cube and then that image was subtracted to each slice. The SDI-processed 3-D spectral cubes were then demagnified and injected into the ADI image subtraction step. We made use of the basic ADI or classic ADI (*65*), LOCI (*73*) with the typical parameters dr=5, Na=200, g=2, and N_delta=0.75 FWHM, and PCA keeping 10% of the modes.

Pipeline 2 follows the approach described in Pueyo et al. (*74*) and uses the KLIP algorithm. For each slice in each exposure the field of view is first divided in subtraction zones defined by Nr annuli and Ns azimuthal sectors. A PSF library is then created for each zone. It is composed of the Ncor most correlated sub-images chosen among the ensemble of slices for which a possible point source located in the center of the zone would have moved by at least N_delta pixels either in the radial (SDI) or azimuthal direction (ADI). The KLIP algorithm is then used in conjunction with this local speckle noise reference library. A search for optimal algorithm parameters was then carried out for each dataset. Typical values are Nr = 7, Ns = 4, N_delta = 3.5 pixels, Ncor/Kklip = 100/10.

In the third pipeline, the TLOCI algorithm (*75*) is combining both SDI and ADI data into a single stellar halo subtraction step. A generic input spectrum was used to guide the reference image selection process to minimize the self-subtraction and maximize the signal-to-noise ratio (S/N) of any companion having a spectrum similar to the input spectrum. A maximum flux contamination ratio of 90% inside a 1.5 lambda/D aperture was chosen for the planet detection mode. The algorithm uses a pixel mask scheme to avoid the planet flux to be fitted by the algorithm (*76*). The least-squares is also using positive coefficients (using positive coefficients avoids large positive and negative coefficients, ensuring that the requested self-subtraction contamination limit is respected) and circular annulus subtraction regions having 1.7 $\lambda/D$ width. The least-squares optimization regions are also circular annulus just inside and outside the subtraction regions, having 3 $\lambda/D$ width for the inner annulus and 6.6 $\lambda/D$ width for the outer annulus.

For all three scripts, the subtracted data cube slices were demagnified, rotated to align their north axis along the image Y-axis, and then median-combined wavelength per wavelength. The last step consists of collapsing the final median-combined subtracted data cube into a broadband final residual image for planet detection (simple addition of the slices (1), weighted mean of the slices (2 and 3) with the coefficients computed from the input spectrum to maximize the S/N).

The NIRC2 *Lp* data were processed using the custom SOSIE IDL script (*76*) that shares many similarities with the TLOCI code (pipeline 3) discussed above.

**Relative astrometry, photometry, and spectral/photometry extraction**: All three pipelines and the automatic GPI pipeline detected 51 Eri b. We measured its signal-to-noise (S/N) to estimate the confidence level of the detection at each epoch. We computed the ratio of 51 Eri b's integrated flux within a circular aperture of 1.5 $\lambda/D$ radius to the noise measured by the standard deviation within an annulus of 1.5 $\lambda/D$ with angular separation of 51 Eri b. No correction for small sample statistics was necessary because the object is at sufficiently wide separation (> 10 $\lambda/D$) that this effect is negligible (*77*).

To perform a detailed astrometric and spectroscopic characterization of 51 Eri b, Pipeline 1 and 3 were rerun using ADI-only data for the stellar halo subtraction to avoid complicated biases when using SDI for spectral extraction. We followed the approach of Lagrange et al. (*3*) and Marois et al. (*76*) by injecting artificial planets. For each slice, the PSF used to create the artificial planets was derived from the average of the four satellite spots averaged over the entire observing sequence. A first guess of the position was used to create a template of the ADI-subtracted point source by applying the same ADI subtraction parameters. A rough spectrum was then obtained at that location by iterating the template point source flux (using positive and negative values) to minimize the difference between the detected point source and the template planet image. The data cube was then collapsed into a broadband, high S/N image to derive the point source position (by again minimizing the residuals after the template point source subtraction). The final point source flux at each wavelength is then extracted using the same iterating technique. The flux and astrometry minimization was performed in a 2×3 $\lambda/D$ area. The measurement errors were computed by performing the same extraction on artificial point sources having the same extracted flux as 51 Eri b and at same angular separation relative to the star, but at different position angles. The photometric and astrometric errors are the standard deviation of the recovered values. Pipeline 2 used both ADI and SDI in conjunction with the KLIP algorithm for spectral characterization, correcting for over-subtraction using the method described in Pueyo et al. (*74*). SDI biases are mitigated using this approach but might not be completely cancelled (depending on algorithm parameters), since some fraction of the point source's flux remains in the PSF library. We ran the extraction over a large grid of parameters (~2000) for both the point source and a series of synthetic point sources at same angular separation as 51 Eri b but different position angles. Optimal extraction parameters (~200) are then inferred by applying to these synthetic sources the astrometric and spectral stability diagnostics detailed in Pueyo et al. (*74*). Final astrophysical estimates on 51 Eri b are obtained then using this well-behaved subset of parameters. Measurement errors are computed as in pipelines 1 and 3.

The broadband star-to-planet relative photometry was computed by adding the flux measured in individual slices and dividing it by the spot intensity. We then used the spot-to-star mean contrast measured during laboratory testing (9.23±0.06 mag for *H*-band and 9.36±0.06 mag for *J*-band, *79*) to obtain the final star-to-planet contrast. The final contrasts and positions, along with their associated errors, measured at *H*-band and *J*-band were computed from the average of the measurements and errors of the three scripts. The final errors on the position and flux were computed by adding quadratically the errors on the planet measurements, the image registration with the four spots (0.05 px), the plate scale (0.014 mas/px *80*), and the north position angle for the former (0.25 deg using astrometric binaries), and on the measurement, the star-to-spot ratio (0.06 mag, *79*), and the spot variability (3% at *H* band, 10% at *J* band) for the later. The final results can be found in Table S2.

To retrieve the spectrum of 51 Eri b, the contrast at each wavelength was multiplied by a template spectrum of the host star. We used an F0-2IV spectrum from the Pickles (*S37*) library and the 2MASS *J* and *H* magnitudes of the star (*81*). Corrections between 2MASS and GPI magnitudes in each filter were computed with the transmission of the filters (from http://www.ipac.caltech.edu/2mass/releases/allsky/doc/sec3_1b1.html#s6) and an absolute flux-calibrated modeled-spectra of Vega (*82*). They were found to be negligible. The final spectra are shown in Figure S2.

The NIRC2 *Lp* photometry was derived using similar techniques. A template ADI-subtracted PSF was generated and used to estimate the planet's position and flux by minimizing the noise after the template PSF subtraction at the point source position. Artificial point sources having the same intensity and at the same separation but different position angles were then used to derive the astrometric and photometric error bars. The star-to-planet contrast was estimated using unsaturated and unocculted 51 Eri images acquired before the coronagraphic sequence. A conservative 1 pixel (10 mas) error was assumed.

**Significance of the detection and analysis**: In the *H*-band December 2014 dataset, 51 Eri b is detected with a S/N of 8-9 in the final collapsed residual image, the brightest signal at that separation (see Figure 1). All three scripts and the automated system were able to detect the source with similar S/N and over a broad range of reduction parameters. The source was also clearly detected in the individual spectral channels. SDI was not necessary to reveal the planet or its methane absorption, but when applied significantly enhances the S/N of the detection. At *J* band, 51 Eri b is not detected in the 2015/01/29 dataset, but it is recovered in the 2015/01/30 dataset by the three scripts at low 3-4 S/N. This detection is much more sensitive to algorithm fine tuning. The *H*-band January data also reveals 51 Eri b at S/N of 6 with SDI. Finally, 51 Eri b is detected at S/N of 5 in the NIRC2 *Lp* dataset. The detection of 51 Eri b is therefore robust, having been found in multiple bands, multiple epochs and with two different telescopes and instruments. Figure S1 shows the *J*-band reduction of the 2015/01/30 dataset by the three pipelines and the H band reduction of the January dataset.

At each epoch, the astrometry and broadband photometry of 51 Eri b derived by the three scripts are consistent within 1-σ or better. The contrast at *H* band between the two epochs is also consistent. However, we only performed full spectral extraction on the first epoch since it has the highest S/N. The three extracted *H*-band spectra are shown in Figure S2. At *J*-band, the lower S/N of 51 Eri b, the lower S/N of the four calibration spots and the more aggressive ADI processing required to see the planet are generating higher discrepancies between the extracted spectra. Intensive subtraction parameter explorations and tests were carried to remove those differences without succeeding. We also injected simulated planets at different position angles with a similar spectrum; all three scripts were able to retrieve the injected spectrum within 5-10%. Our discrepancies likely come from a the presence of a negative speckle, close to and of the same intensity as 51 Eri b that the pipelines are treating differently, causing different flux estimates. Alternatively the azimuthal heterogeneities in speckle statistics (associated with AO residuals) complicate the tests using synthetic sources. In spite of these differences all three pipelines retrieve the peaky features of the *J*-band spectrum.

**Archival data**: We retrieved observations of the star taken with CIAO/Subaru at *K* band in 2003 (PI: Liu), NIRC2/Keck at *H* band in 2003 (PI: Liu), at *Lp* band in 2011 (PI: Kalas), NaCo/VLT at *Lp* in 2009 ( PI: Rameau, with a non-detection published in *84*), and NICI/Gemini at CH4 (PI: Liu, with a non-detection published in *35*). These data were reduced in a similar way as the GPI data and showed no evidence of a point source at the expected location of 51 Eri b, due to a lack of sensitivity close to the parent star. The most sensitive dataset is the 2003 NIRC2/Keck data but they still would not have detected the planet at its current separation of 450 mas. A point source of the *H*-band brightness measured by GPI would have been detectable outside of 1300 mas.

**Astrometry Analysis**: The different GPI observations are separated by only 44 days. If 51 Eri b was a stationary background object seen in the *H*-band December 2014 data, it would have moved by about 14 mas (1 mas in separation, 1.8 deg in position angle) by the end of January. However, the stationary background object scenario can be ruled out given the non-detection of the candidate in the 2003 NIRC2/Keck data, since the parallactic and proper motion of the star 51 Eri would mean that in 2003 the candidate would be at a separation of 1.4", where it would be detectable, but is not observed. Were 51 Eri b instead a field brown dwarf, unbound to 51 Eri, it would have non-zero proper motion and parallax, which could possibly be oriented to be consistent with the astrometry from December 2014 to February 2015 and also the non-detection in 2003. Such a chance alignment of an unbound T-dwarf is disfavored statistically as we detail below. The 2003 non-detection and the 2014-2015 detections constrains the proper motion of 51 Eri b in RA to between -37 and 149 mas/yr and in DEC between -121 to -2 mas/yr at 95% confidence, excluding a stationary background object but of course consistent with the (43, -64) mas/yr proper motion of 51 Eri itself.

Additional epochs of astrometry in late 2015 when the star is observable again will allow us to rule out additional combinations of proper motion and parallax and potentially detect orbital motion, depending on the orientation of the orbit.

**Atmosphere Modeling:** 51 Eri b has a spectrum unmistakably shaped by methane absorption and, as such, is clearly an object in the substellar mass regime. To gain better insight into the properties of this object, we turn to atmosphere models that include the relevant low-temperature atmospheric physics and chemistry. These models have been extensively tested by observations of brown dwarfs and exoplanets with varying degrees of success (*11*, *12*, *84*). The spectral and photometric observations were compared to two sets of synthetic model spectra, produced independently by members of the GPIES team using distinct atmosphere codes (*85*, *86*).

Fitting cloudless, solar metallicity, chemical equilibrium atmosphere models (36) to the *J*- and *H*-band spectra yields good agreement to the observations (Figure 2); however, the best-fitting model has parameters beyond the edge of the model grid, with $T_{\mathrm{eff}} \sim 750$ K and log(*g*) > 5.5 (cgs units). Such solutions imply ages far in excess of the age of the system (well over 5 billion years). A similar fitting procedure was repeated, but for a second model grid with similar assumptions but produced with a different code (12). The best matches to the complete dataset using this second grid of models have $T_{\mathrm{eff}}$ between 900 and 1200K. A similar temperature range is found when only the *H*-band spectrum is fit. In agreement with the first model grid comparison, the surface gravity is not well constrained by the data, and acceptable fits were found with log(*g*) between 2.5 and 5.5 (cgs units). From these model fits, the bolometric luminosity was found to be log $L_{\mathrm{bol}}/L_{\odot} = -5.6 \pm 0.2$. Bolometric corrections determined for brown dwarfs (*S45*) with similar spectral shape as 51 Eri b yield $L_{\mathrm{bol}}$ that agree within our uncertainties.

Model solutions with temperatures > 1000 K equate to small radii (< 0.5 $R_{\mathrm{J}}$) for the inferred bolometric luminosity. From our understanding of brown dwarfs, transiting exoplanets, and the giant planets of our solar system, we expect that all hydrogen/helium dominated objects should have radii close to, or greater than, that of Jupiter. We can, therefore, reject the high-temperature cloud-free fits, even though some have small chi-squares. A few high surface gravity atmosphere models are also able to match the observed fluxes reasonably well but are also discarded for reasons discussed above.

We further focus our parameter search by using the inferred $L_{\mathrm{bol}}$ above and models that describe the cooling history (evolution of $T_{\mathrm{eff}}$, radius) of substellar objects (*36)*. Even though at the earliest ages (< ~ $10^8$ yrs) the cooling histories are most uncertain, the range of possible initial conditions (set by the formation processes) are well constrained by theory (*15, 18*). If we bias the atmosphere model fits to allow only $T_{\mathrm{eff}}$ and gravity that fall within the range predicted by cooling tracks, we find that $T_{\mathrm{eff}}$ is between 550 and 750K (but still no strong constraint on gravity). We adopt this temperature range for 51 Eri b.

Assumptions of solar-abundances, equilibrium chemistry, and cloud-free are merely matters of convenience, greatly reducing the number of free parameters for the model atmosphere grids. Clouds are important sources of opacity in previously discovered directly imaged planets (*1)* and recent evidence suggests that clouds may play an important role even in low-temperature brown dwarfs previously thought to be completely cloud free (*88*). Deviations from equilibrium chemistry are also now considered important in all substellar objects and may delay the onset of methane dominance in near-IR spectra of young giant planets (*11*, *89*, *90*). Consequently, one should consider the model comparison described above as only a first step.

To test the importance of clouds and non-equilibrium chemistry, we explored a small grid of models with varying levels of cloud thickness and vertical mixing of CO to $CH_4$ with an eddy diffusion coefficient, $K_{zz} = 10^8$ cm$^2$ s$^{-1}$. Vertical mixing quenches the mole fractions of $CH_4$ and CO in the atmosphere where chemical and mixing timescale become comparable, resulting in a decrease in $CH_4$ mole fractions in the spectrum-forming region of 51 Eri b's atmosphere. We find that, non-equilibrium chemistry tends to improve the model comparisons across the region of methane absorption (wavelengths between 1.6 and 1.8 microns) for lower gravity more so than for higher gravity atmospheres. Also clouds must diminish quickly above the 10-bar level to avoid reddening the near-IR spectrum too severely. Our best match to the data, with the physical and evolution constraints on radius and effective temperature, was found through a linear combination of a cloudy and cloud-free model spectra (weighted equally) having $T_{\mathrm{eff}}$ = 700K and log ($g$) = 3.5 and including non-equilibrium chemistry (see Figure 2). Combinations of cloudy and cloud-free models have been invoked before in the study of brown dwarfs (*36)* and directly imaged planets (*13, 15, 40*). However, cloud properties, gravity and metallicity all impact the overall spectral shape, often in a similar manner (*91*), making it difficult to uniquely explore all possible combinations of atmospheric ingredients and cloud-covering fractions. Furthermore, while our two-component model comes close to matching the observations as a whole, it does not represent a self-consistent atmosphere. The two models that have been combined have different temperatures and pressures in their deeper convective regions – as such, they do not ideally represent two different regions of the same object because their temperature profiles do not converge upon the same interior temperature and pressure. A two-column model (*14)* that enforces consistency within the adiabatic region would be an improvement, but a grid of such models is not yet available for a comparison. Until then, we put forth our two-component model as merely suggestive that 51 Eri b is best represented by a low-gravity atmosphere with $T_{\mathrm{eff}} \sim$ 700 K and a non-homogeneous contribution from cloud opacity across its surface.

**Calculation of background T dwarf contamination rates**

Computation of the detection rate of physically unassociated, methane-dominated, field T dwarfs requires knowledge of their volume density and their luminosity. As we are probing only the local solar neighborhood (< 150 pc) we assume that their distribution is isotropic and neglect the decrease of volume density with increasing distance above the Galactic plane. As such, this estimate will overestimate the contamination rate.

The luminosity function of T dwarfs is measured using wide field optical and infrared surveys. No single survey is sensitive to the entire spectral range; therefore, we have combined the results of Reyle et al. (*92*) for T0-T5.5 and Burningham et al. (*93*) for the later spectral types. The adopted luminosity function is listed in Table S3.

To find the false alarm rate we calculate the distance, $d_5$, out to which each T spectral type yields a detection of significance ≥ 5-σ, compute the corresponding volume, and multiply by the volume density from Table S3. We ignore binaries, the luminosity dispersion within each spectral bin, and compute the mean absolute *H*-band magnitudes for each spectral type from the data in Kirkpatrick et al. (*34*). We then ask how many T-class objects are expected in the GPI field of view by assuming a uniform sensitivity in an annular field between radii of 0.2 and 1 arc second. Performing the integral over the luminosity function for detections ≥ 5-σ and all types ≥ T0 yields $1.72 \times 10^{-7}$ T-type objects per GPI field. The proper motion constraint from the 2003 Keck non-detection further eliminates 66% of background objects for typical T-dwarf proper motions. Combining the false alarm rates of all 44 fields surveyed to date with GPI produces a total false alarm rate as of January 2015 of $2.4 \times 10^{-6}$ methane-dominated non-companions.

**SOM Figures**

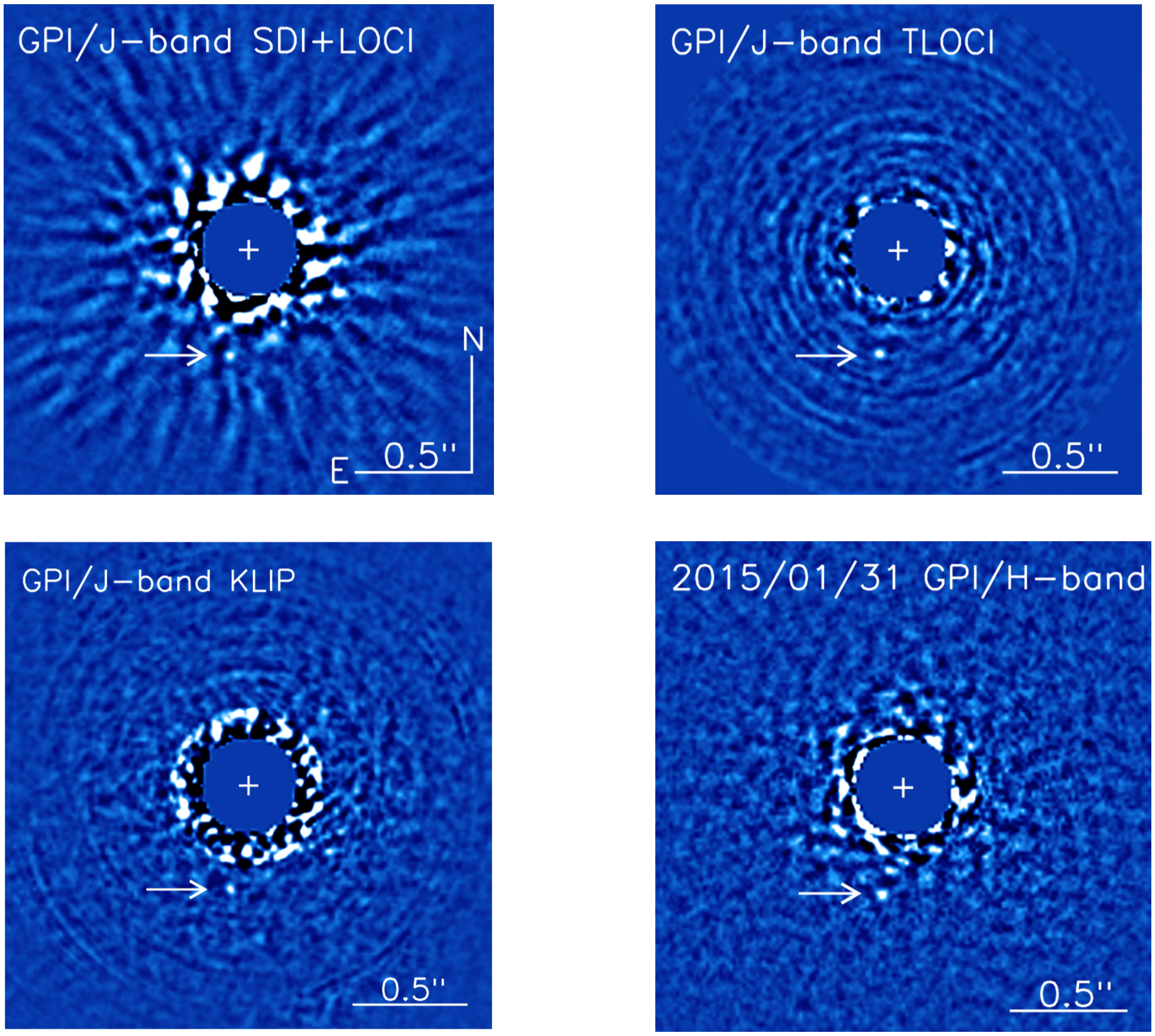

**Figure S1: Residual images showing 51 Eri b at *J* and *H* bands from the 2015 January data. A negative speckle is seen at *J* band north east of 51 Eri b. First panel: *J*-band SDI+LOCI reduction (1) with dr = 5, Na = 200, g = 2, and N_delta = 0.75 FWHM. Second panel: *J*-band TLOCI reduction (3) assuming a methane input spectrum. Third panel: *J*-band Klip reduction (2) with Nr = 7, Ns = 12, N_delta = 2.5 pixels, Ncor/Kklip = 200/20. Fourth panel: *H*-band SDI+LOCI (1) with the same parameters as for *J*-band.**

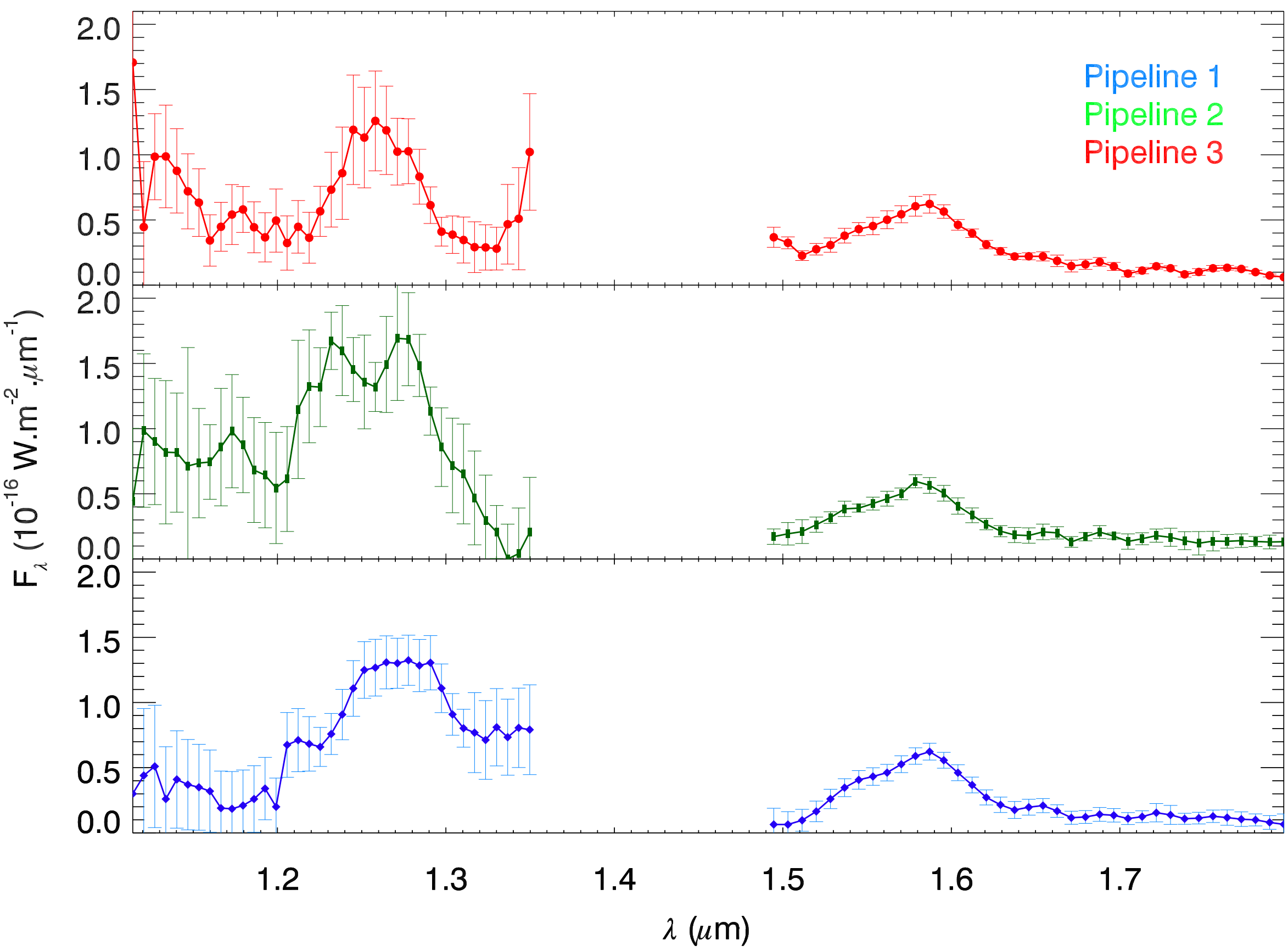

**Figure S2: 51 Eri b spectra from the three different pipelines. The *H*-band spectra agree well while at *J*-band, the discrepancies are explained by the combination of the low S/N and the presence of a close by negative speckle of similar intensity that the pipelines treat differently.**

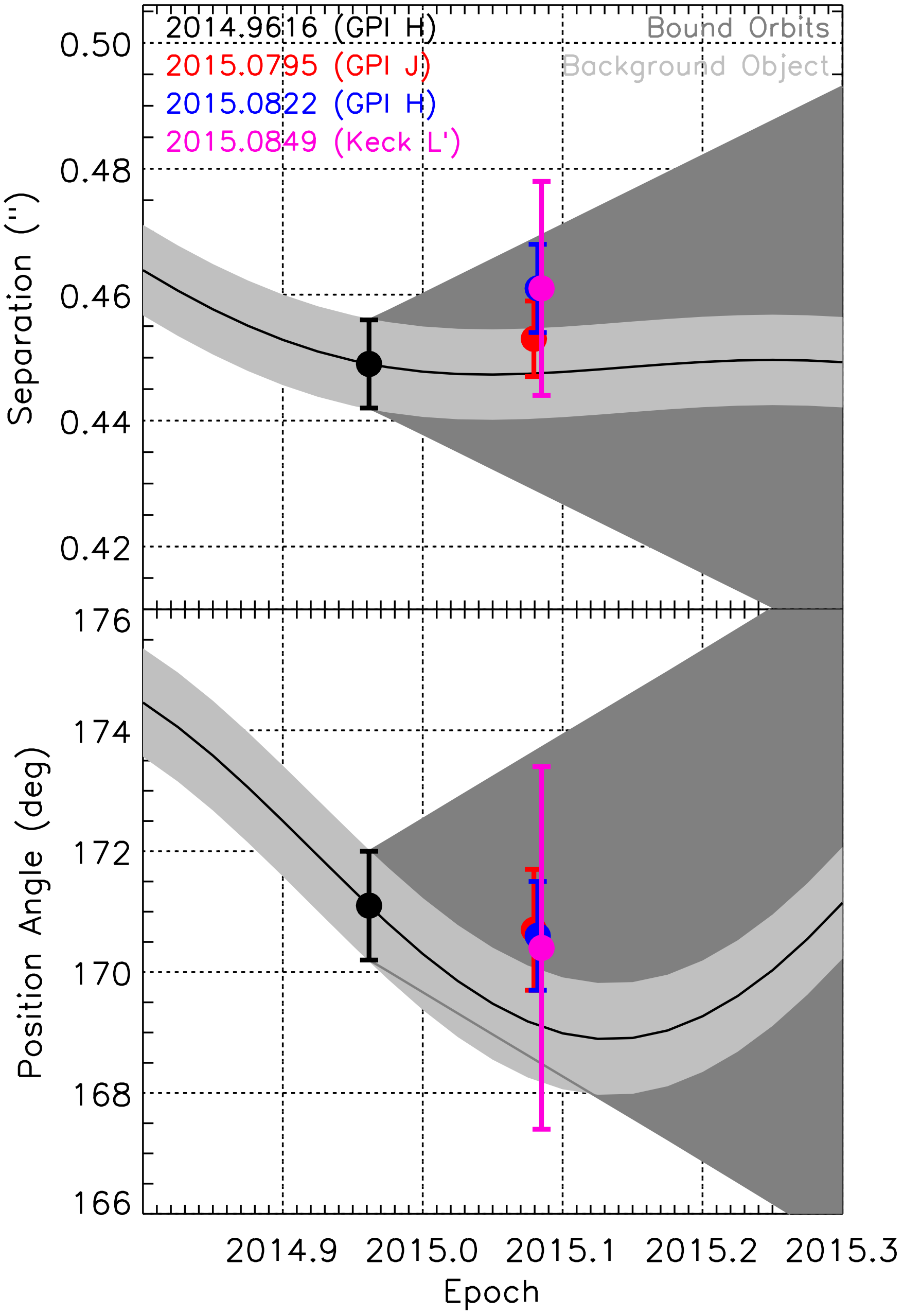

**Figure S3: Astrometry of 51 Eri b. The light gray track indicates the motion of a background object while the dark gray track shows the maximum possible orbital motion (escape velocity for 1.2   M_sun). Both the low precision of the measurements and small proper and parallactic motion of 51 Eri between the epochs cannot assess the astrometric status of the companion.**

**SOM Tables**

**Table S1: 51 Eri observing log**

| Inst./Telescope | UT Date | Filter | Exposure Time (s) | FoV rotation (deg) | Average obs. Condition |
|---|---|---|---|---|---|
| | | | | | |
| GPI/Gemini South | 2014/12/18 | *H* | 2280 | 23.8 | Excellent |
| | 2015/01/29 | *J* | 3120 | 31.8 | Poor |
| | 2015/01/30 | *J* | 4200 | 37.7 | Good |
| | 2015/01/31 | *H* | 3600 | 32.1 | Average |
| NIRC2/Keck | 2015/02/01 | *Lp* | 3348 | 44.6 | Excellent |

**Table S2: Astrometry and photometry of 51 Eri b from all three pipelines. Adopted values are obtained from the mean of the three scripts, except for the *H*-band of January and *L'* dataset where one script was used.**

| UT Date | Filter | Pipeline | Separation (mas) | Position angle (deg) | Contrast (mag) |
|---|---|---|---|---|---|
| 2014/12/18 | *H* | Adopted | 449±7 | 171.1±0.9 | 14.43±0.19 |
| | | 1 | 443±5 | 171.3±0.6 | 14.44±0.13 |
| | | 3 | 453±9 | 170.9±1.1 | 14.46±0.24 |
| | | 2 | -- | -- | 14.41±0.20 |
| 2015/01/30 | *J* | Adopted | 453±6 | 170.7±1.0 | 14.34±0.39 |
| | | 1 | 455±6 | 170.7±1.2 | 14.36±0.26 |
| | | 3 | 450±6 | 170.7±0.8 | 14.17±0.38 |
| | | 2 | -- | -- | 14.50±0.50 |
| 2015/01/31 | *H* | Adopted/1 | 461±7 | 170.6±0.9 | 14.50±0.34 |
| 2015/02/01 | *L'* | Adopted/3 | 461±24 | 170.4±3.0 | 11.62±0.17 |

**Table S3: T-dwarf luminosity function and maximum detection distance.**

| Spectral Type | $M_{\mathrm{H}}$ | Number | $d_5$ |
|---|---|---|---|
| | (mag.) | (pc$^{-3}$/SpT/1 ×10$^{-3}$) | pc |
| T0-0.5 | 13.6 | 0.3 | 165 |
| T1-1.5 | 13.6 | 0.3 | 169 |
| T2-2.5 | 13.6 | 0.3 | 168 |
| T3-3.5 | 13.7 | 0.3 | 160 |
| T4-4.5 | 13.9 | 0.3 | 141 |
| T5-5.5 | 14.4 | 0.3 | 115 |
| T6-6.5 | 15.1 | 0.5 | 85 |
| T7-7.5 | 15.9 | 0.7 | 56 |
| T8-8.5 | 17.0 | 2.7 | 34 |